\documentclass[comsoc, journal]{IEEEtran}
\usepackage[utf8]{inputenc}
\usepackage[T1]{fontenc}

\usepackage{graphicx}
\usepackage{subfig}
\usepackage{float}
\usepackage{enumitem}

\usepackage{array, makecell, multirow, booktabs}
\newcolumntype{P}[1]{>{\centering\arraybackslash}p{#1}}
\setcellgapes{3.5pt}

\makeatletter
\def\ps@IEEEtitlepagestyle{%
  \def\@oddfoot{\mycopyrightnotice}%
  \def\@evenfoot{}%
}
\def\mycopyrightnotice{%
  {\footnotesize  This paper has been accepted for publication by the IEEE Wireless Communications Magazine. The final version will be published by the IEEE. \hfill}
  \gdef\mycopyrightnotice{}
}

\begin{document}

\title{Information-Centric Networking in Wireless Environments: Security Risks and Challenges}

\author{
    Boubakr Nour, Spyridon Mastorakis, Rehmat Ullah, and Nicholas Stergiou

    \thanks{B. Nour (corresponding author) is with Beijing Institute of Technology, Beijing, China (email: n.boubakr@bit.edu.cn).}
    
    \thanks{S. Mastorakis and N. Stergiou are with the University of Nebraska at Omaha, Omaha, US (email: smastorakis@unomaha.edu, nstergiou@unomaha.edu).}
    
    \thanks{R. Ullah is with the School of Electronics, Electrical Engineering and Computer Science, Queen's University Belfast, United Kingdom (email: r.ullah@qub.ac.uk).}
}


\maketitle

\begin{abstract}
	Information-Centric Networking (ICN) has emerged as a paradigm to cope with the lack of built-in security primitives and efficient mechanisms for content distribution of today's Internet. However, deploying ICN in a wireless environment poses a different set of challenges compared to a wired environment, especially when it comes to security. In this paper, we present the security issues that may arise and the attacks that may occur from different points of view when ICN is deployed in wireless environments. The discussed attacks may target both applications and the ICN network itself by exploiting elements of the ICN architecture, such as content names and in-network content caches. Furthermore, we discuss potential solutions to the presented issues and countermeasures to the presented attacks. Finally, we identify future research opportunities and directions.
\end{abstract}

\IEEEpeerreviewmaketitle

\section{Introduction}
The current Internet infrastructure is witnessing an explosive growth in terms of connected devices and the amount of generated content. The retrieval of data from such devices and the exchange of data among these devices face various issues and challenges, such as the efficiency of data delivery and the security of the overall process. The current host-centric Internet architecture was developed several years ago to allow connectivity between devices. However, nowadays, users and applications seek to retrieve content of their interest. To this end, the host-centric Internet architecture cannot adequately fulfill this premise with its current stack and protocols.

Information-Centric Networking (ICN)~\cite{xylomenos2014survey} has been proposed as a future Internet architecture. In contrast to IP-based networks where the communication is based on host addresses, ICN is based on a content-centric communication model that directly targets the content regardless of the addresses of hosts. ICN aims to decouple the content from its hosting location(s), while ICN forwarders cache the content in the network and serve future client requests without forwarding them to the original producer.

\begin{figure}[t]
	\centering
	\includegraphics[width=\linewidth]{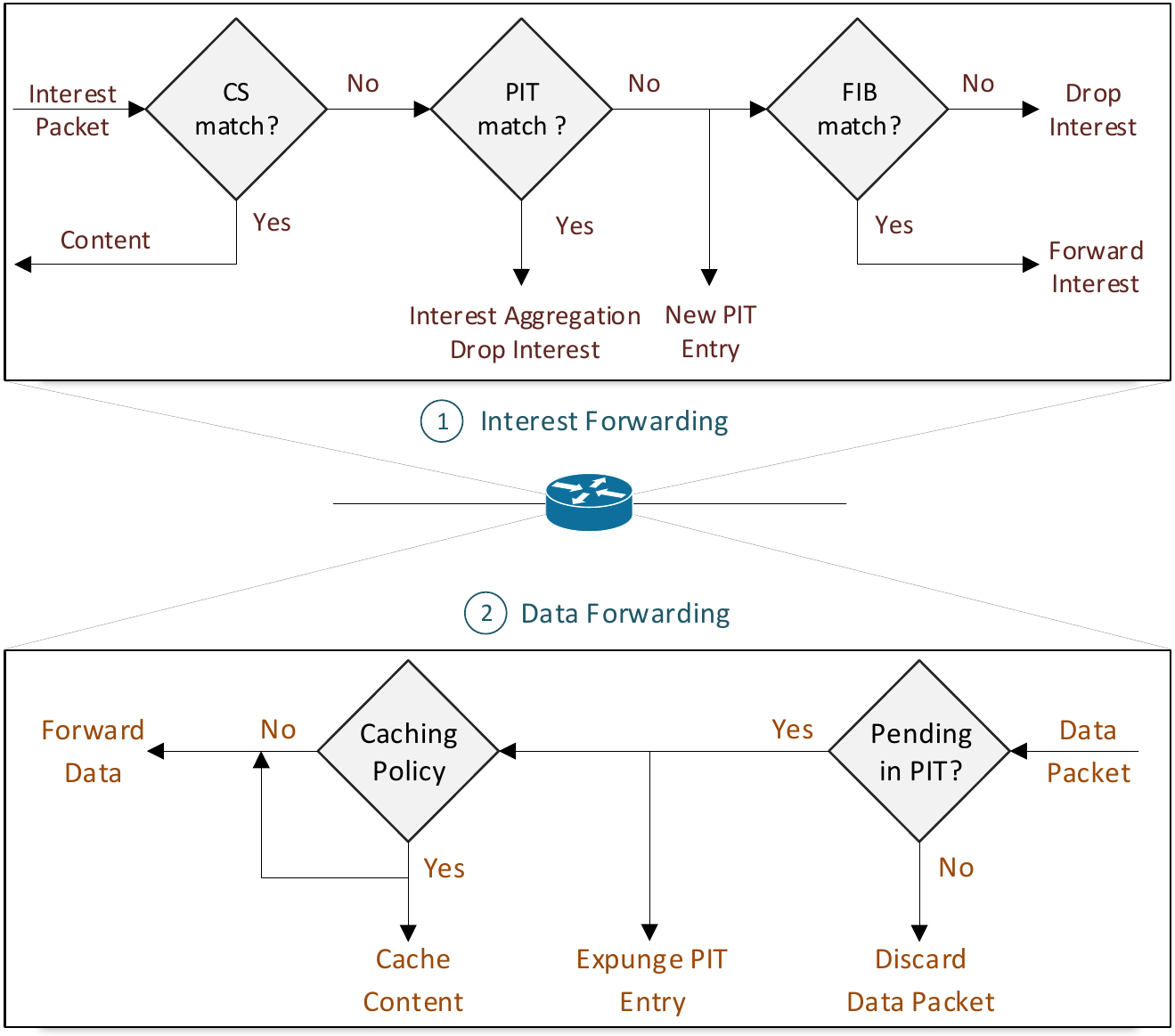}
	\caption{Interest/data forwarding plane.}
	\label{fig:ndn_forwarding}
\end{figure}

In ICN, content discovery and delivery are receiver-driven, as illustrated in Figure~\ref{fig:ndn_forwarding}. A consumer initializes a content request, called an Interest packet, that carries the name of the requested content. This packet is forwarded hop-by-hop using name-based routing until reaching the content producer (original data producer or a replica intermediate node that has the requested content). Subsequently, a response (Data packet) is delivered back to the requester(s) using the reverse Interest path. In contrast to host-based networks where securing the channel between hosts is vital, ICN adopts a content-centric security model that directly secures the content regardless of the communication channel~\cite{yu2018content}.

By naming and securing data directly at the network layer, essentially decoupling the content from its original producer, ICN is a promising architecture for wireless networks. ICN offers native support for mobility, data multicast, and ubiquitous in-network content caching. At the same time, the content can be served from any node that can offer it and can be consumed by users in an asynchronous manner without requiring the producer to be constantly available. 

Inspired by the use case of content distribution on the Internet, the majority of ICN research so far has focused on wired rather than wireless networks (either infrastructure-based or infrastructure-less). Deploying ICN in wireless environments may result in several security issues stemming from both the nature of wireless communication itself and the content-centric communication model of ICN. Such issues may include attacks on content names that affect user privacy, managing access control rules that involve content names, establishing trust in wireless environments, and authenticating data delivered from content stores, among others~\cite{zhang2018overview}.

In this paper, we focus on security risks and challenges that may arise when ICN is deployed in wireless networks. We make the following contributions: (i) we present several security issues and attacks against the architectural features of ICN in wireless environments; and (ii) we discuss potential solutions to these issues and identify directions for future research. The rest of this paper is organized as follows: In Sections~\ref{sec:principles} and~\ref{sec:wireless}, we summarize the ICN design principles and explain how ICN can benefit wireless communications respectively. In Section~\ref{sec:issues}, we present security issues in wireless ICN networks and potential solutions. Finally, in Section~\ref{sec:conclusion}, we conclude our paper and summarize future research directions identified throughout the paper.

\section{ICN Design Principles}
\label{sec:principles}
Among the various ICN implementations that have been proposed in the literature, Named Data Networking (NDN)~\cite{zhang2014named} is one of the recent and active ICN realizations. NDN is a pull-based architecture that uses an Interest-Data communication model. In NDN, a requester sends an Interest packet asking for content by name. Interest packets are forwarded upstream (towards data producers) through name-based routing.

\textbf{NDN Forwarding Design:}
During the Interest forwarding process (Figure~\ref{fig:ndn_forwarding}), intermediate forwarder nodes check their \textit{Content Store} (CS) for existing cached data. If a match is found, they reply with the data immediately without forwarding the received Interest to the content producer. Otherwise, a lookup in the \textit{Pending Interest Table} (PIT) is performed. If an Interest with the same name is currently pending in PIT, the received Interest is dropped, and the incoming interface of the Interest is appended to the found PIT entry, a concept called \textit{Interest aggregation}. Otherwise, a lookup in the \textit{Forwarding Information Table} (FIB) is performed to identify next-hops after creating a new PIT entry for the Interest.
During data delivery, intermediate forwarders check their PIT for each received Data packet. If a match is found, the data is forwarded to all interfaces listed in the corresponding PIT entry (\textit{data multicast}). If a match is not found in PIT, the packet is considered unsolicited and is dropped.

\section{NDN/ICN in Wireless Networks}
\label{sec:wireless}
\begin{figure*}[t]
	\centering
	\includegraphics[width=\linewidth]{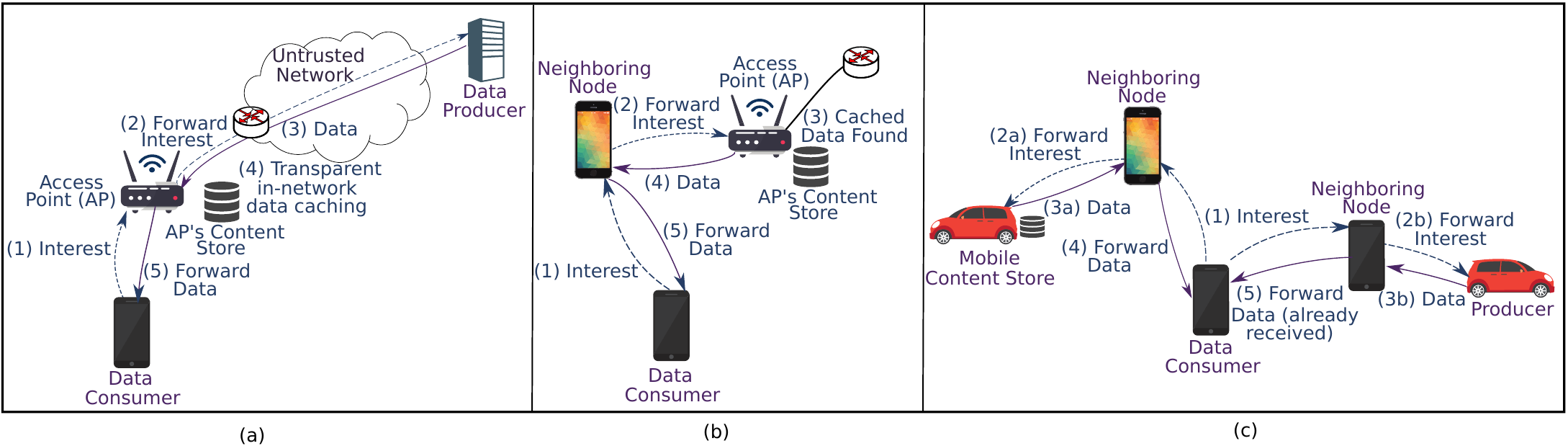}
	\caption{(a)~Content retrieval from a producer on the Internet, (b)~Cached content response from an AP, (c)~Content response from neighboring nodes.}
	\label{fig:commModel}
\end{figure*}

Various NDN/ICN features can inherently support and enhance the nature of wireless communication. Let us consider an infrastructure-based scenario where $N$ wireless devices are connected to the same Access Point (AP) and are downloading the same content. At least $N + 1$ virtual unicast channels must be created within the broadcast channel to deliver the content to all users. On the contrary, the ICN paradigm can take advantage of content names and the inherently broadcast nature of wireless channels to deliver the content to all users through one virtual channel. Similarly, if the requests are disjoint, the use of in-network caching (e.g., by the AP) may drastically reduce the amount of content carried on the backhaul link and enhance the bandwidth and energy consumption, since all requests are satisfied from the AP's CS. Figure~\ref{fig:commModel} illustrates different Interest-Data (request-response) exchange patterns in a wireless network. The communication could be performed through an AP or in an ad-hoc manner. A request from a wireless node can be satisfied either by the original producer, any CS in the communication path (e.g., at the AP or an edge server), or devices within the requester's communication range (e.g., vehicles in a vehicle-to-vehicle scenario).

NDN/ICN can benefit wireless communications in the following ways: 

\begin{itemize} [leftmargin=0cm,itemindent=.3cm,labelwidth=\itemindent,labelsep=0cm,align=left, noitemsep, topsep=0pt]
	\item \textit{Clean architecture - less overhead:} The clean design of NDN provides a simple, but effective request-response communication model for wireless networks initiated by data requesters, without involving host addresses.
	
	\item \textit{Localized content retrieval/sharing:} The name-based communication model enables localized content discovery and retrieval from any entity that can provide the requested content across a single hop or more wireless hops away from the requester.
	
	\item \textit{Content distribution scalability:} The use of content names and the receiver-driven communication model enhance the scalability of the content distribution process as the number of devices or the amount of generated content increases.
	
	\item \textit{Strategic in-network caching points:} Caching the content in the network based on the capabilities of each forwarder helps the network deliver the content without requiring the wireless devices to be continuously connected. To this end, in-network caching enhances data availability, dissemination, and decreases the data retrieval latency.
	
	\item \textit{Simplified deployment (underlay \& overlay):} Through content names, the deployment of wireless applications is simplified. Applications specify only the content name without having to know the address of the content producer. NDN may also be deployed as an overlay on top of TCP/IP, or as a clean-slate solution replacing TCP/IP.
\end{itemize}

In the host-centric network paradigm, all devices are required to forward their requests towards the AP that acts as a connectivity bridge between devices and the Internet. Even if the destination device resides within the local network, the request must be handled by the AP. However, in NDN/ICN, the AP may cache popular content to satisfy incoming consumer requests. More importantly, due to the use of content names, NDN/ICN enables machine-to-machine communication without the need to involve the AP in the communication process.

\section{Security Issues in Wireless NDN/ICN}
\label{sec:issues}
Moving away from host addresses to content names changes how security mechanisms are realized. In ICN, security becomes a property of the data, which secures the data directly instead of securing the communication channel. However, this does not imply that communication channels become insecure, but it rather opens new research avenues for channel/path/session management. Table~\ref{tab:icn_attacks} provides a list of attacks in ICN. These attacks can take place in both wired and wireless environments. The remainder of this section focuses on attacks that can occur in ICN-based wireless environments.

\begin{table*}[t]
	\centering
	\makegapedcells
	\caption{A sample list of security attacks and vulnerabilities in NDN/ICN.}
	\label{tab:icn_attacks}
	\begin{tabular}{|l|l|l|l|l|}
		\hline
		\multicolumn{2}{|c|}{\textbf{Attack}} &
		\multicolumn{3}{c|}{\textbf{Objective}} \\ \hline
		
		\multicolumn{1}{|c|}{\textit{ICN Component}} &
		\multicolumn{1}{p{0.16\textwidth}|}{\textit{Attack Type}} &
		\multicolumn{1}{c|}{\textit{Launched by}} &
		\multicolumn{1}{c|}{\textit{Target Layer}} &
		\multicolumn{1}{c|}{\textit{Security Implications}} \\ \hline
		
		\multicolumn{1}{|c|}{\multirow{3}{*}{Content Name}} &
		\multicolumn{1}{p{0.16\textwidth}|}{Interest Flooding Attack} &
		\multicolumn{1}{l|}{Consumer} &
		\multicolumn{1}{l|}{NDN} &
		\multicolumn{1}{p{0.35\textwidth}|}{
			$\bullet$ Degrades network/service performance. \newline
			$\bullet$ Degrades data/content availability.
		} \\ \cline{2-5}
		
		~ &
		\multicolumn{1}{p{0.16\textwidth}|}{Monitoring Attack} &
		\multicolumn{1}{l|}{Passive Listener} &
		\multicolumn{1}{l|}{Application} &
		\multicolumn{1}{p{0.35\textwidth}|}{$\bullet$ Monitors the user requests. \newline 
			$\bullet$ Surveillance of popular content. \newline
			$\bullet$ Violates the user privacy.
		} \\ \cline{2-5}
		
		~ &
		\multicolumn{1}{p{0.16\textwidth}|}{Watchlist Attack} &
		\multicolumn{1}{l|}{Forwarder} &
		\multicolumn{1}{l|}{NDN} &
		\multicolumn{1}{p{0.35\textwidth}|}{$\bullet$ Filters or blocks a predefined list of content names.
		} \\ \hline
		
		\multicolumn{1}{|c|}{\multirow{2}{*}{Content Caching}} &
		\multicolumn{1}{p{0.16\textwidth}|}{Content Poisoning Attack} &
		\multicolumn{1}{l|}{Any Node} &
		\multicolumn{1}{l|}{NDN} &
		\multicolumn{1}{p{0.35\textwidth}|}{$\bullet$ Alters data availability and correctness.} \\ \cline{2-5}
		
		~ &
		\multicolumn{1}{p{0.16\textwidth}|}{Cache Pollution Attack} &
		\multicolumn{1}{l|}{Consumer} &
		\multicolumn{1}{l|}{NDN} &
		\multicolumn{1}{p{0.35\textwidth}|}{$\bullet$ Changes the data popularity and distribution. \newline 
			$\bullet$ Eliminates the utility of in-network caching. \newline 
			$\bullet$ Increases the request latency and network load.
		} \\ \hline
		
		\multicolumn{1}{|c|}{\multirow{2}{*}{Data/Content}} &
		\multicolumn{1}{p{0.16\textwidth}|}{Timing Attack \newline Mistreating Attack} &
		\multicolumn{1}{l|}{Passive Listener} &
		\multicolumn{1}{l|}{NDN/Application} &
		\multicolumn{1}{p{0.35\textwidth}|}{$\bullet$ Spoofs the routing/forwarding information. \newline
			$\bullet$ Violates the user and data privacy. \newline
			$\bullet$ Manipulates content integrity and correctness.
		} \\ \cline{2-5}
		
		~ &
		\multicolumn{1}{p{0.16\textwidth}|}{Unauthorized Access} &
		\multicolumn{1}{l|}{Any node} &
		\multicolumn{1}{l|}{Application} &
		\multicolumn{1}{p{0.35\textwidth}|}{$\bullet$ Breaks access control rules. \newline 
			$\bullet$ Gains access to privileged content.
		} \\ \hline
		
	\end{tabular}
\end{table*}

\subsection{NDN Design Issues}
Any node in wireless ICN networks (e.g., content consumer or passive listener who receives packets due to the broadcast nature of the wireless channel without requesting them) can be malicious and exploit features of the NDN architecture to launch various attacks. In combination with the existence of a single wireless (inter)face per device and the content-centric communication nature of ICN, wireless communication in broadcast mode (e.g., in infrastructure-less environments) is facilitated in ICN, while unicast communication faces challenges. However, broadcast wireless communication makes unclear how to utilize NDN's Interest aggregation and whether PIT would be effective overall. At the same time, nodes that overhear others' transmissions can promiscuously cache and serve content without informing the sender or the original content producer. We further discuss these critical issues below, presenting solutions and directions for future research.

\vspace{0.2cm}
\textbf{Single Wireless (Inter)Face:}
In NDN, the term \textit{faces} describes both physical (wired or wireless) interfaces and application logical interfaces to the NDN layer for sending/receiving packets. During the Interest forwarding process in wireless environments, a node has only one wireless interface. As a result, the Interest is passively broadcast at the data link layer (we refer to this mechanism as blind forwarding), since the corresponding PIT entry has only one face listed regardless of the number of requests received. This creates an anomaly for Interest aggregation, as the node cannot differentiate whether the Interest is coming from the same or a different consumer. Similarly, during data forwarding, the Data packets are broadcast in the wireless channel without knowing who has requested the content.

In existing systems, nodes within the sender's wireless range receive transmitted packets but discard the ones that do not contain their physical address as a destination, such as a Media Access Control (MAC) address. However, in NDN, it is yet to be seen if a new, data-centric MAC layer may be standardized or if a modification of the existing MAC layer will be utilized. Another aspect that is yet to be seen is whether NDN wireless nodes will operate by default in promiscuous mode at the link layer, and if so, at which layer packet filtering may take place.

In a wired network, an intermediate node can distinguish among requesters based on which face the Interest has been received from and record this information in PIT in order to use it later for data forwarding. However, in wireless environments, the use of a single interface by intermediate nodes may jeopardize the multicast nature of NDN. Assuming that we utilize the broadcast nature of the wireless medium, each Interest will be broadcast from one node to another. The challenge will be for nodes to decide when to suppress the transmission of a received Interest, thus avoid broadcasting this Interest further. A naive approach might be for nodes to employ a random timer before broadcasting an Interest, essentially overhearing transmissions of the same Interest from their neighbors. If another node broadcasts the Interest first, neighboring nodes will suppress the transmission of the same Interest. Further research is needed towards designing sophisticated Interest forwarding/suppression algorithms.

Approaches that take advantage of mappings between specific MAC addresses and content names have been proposed to achieve unicast transmission at the data link layer through existing MAC protocols~\cite{kietzmann2017need}. Such approaches can effectively mitigate the issues that arise from having a single wireless face per device and deserve further investigation. Finally, designing a new named, content-centric MAC layer protocol matches the ICN philosophy and should be further explored as well.

\vspace{0.2cm}
\textbf{Interest Aggregation:}
The explicit use of broadcast transmissions combined with content names during Interest and data forwarding in wireless environments makes the utility of Interest aggregation unclear. Although the nodes will not forward an Interest which has already been processed (but not satisfied), the nodes may broadcast an Interest at least once. As discussed earlier, this removes the multicast benefit from the wireless domain, and only after the request reaches an AP, some sort of hierarchy can be formed.

A proposed solution may be the use of a \textit{mapping mechanism} to enforce that only the access point will receive the Interests ignoring other intermediate nodes. However, this will neglect the cache utility and lead the network design back to a host-oriented philosophy.

\vspace{0.2cm}
\textbf{PIT Effectiveness:}
In light of the previous discussion, another issue arises with PIT effectiveness. When a malicious node receives a request, it can generate a bogus Data packet with a valid content name as a response to the received Interest. As the Data packet is also broadcast back to the consumer, the AP (or a forwarder node) will receive the packet, satisfy the associated PIT entry, and forward the data towards the consumer. Once the legitimate data is sent in response to the Interest, it will be discarded, as no corresponding PIT entry will exist at intermediate nodes. Work in~\cite{sivaraman2020optimal} studies the optimal PIT size under various network conditions. The study found that the number (distribution) of Interest packets affects the dropping probability while the sojourn time distribution also impacts the node performance.

A hop-by-hop signature verification mechanism can be used to solve this problem. However, performing such a task at intermediate nodes results in additional overhead and communication delay. On the other hand, encrypting the content name may further complicate the communication process because intermediate nodes will need keys to decrypt the received packet names.

\vspace{0.2cm}
\textbf{Illegal Content Caching:}
The in-network caching process is transparent to consumers and producers. Any node can cache the content without informing or receiving permission from the original producer and reply to future Interests without informing the consumer that the content has been served from a cache. However, this may not always be desirable. A producer may not want its content to be cached for several reasons (e.g., copyrights, financial concerns).

Contrary to wired networks where devices can be closely monitored (especially at the network edge and core), in a wireless environment, any node can become part of the network and cache data. Such a node may modify the data and distribute it as original. The consumer can determine if the data is authentic, but there is no guarantee that the consumer will eventually retrieve legitimate data.
Ensuring trusted data caches and avoiding caches that provide bogus data are promising research directions.
Work in~\cite{khelifi2020blockchain} adopts Blockchain technology to secure data forwarding and caching. The work uses a reputation-based system to prevent forwarding malicious/invalid requests or caching bogus data.

New business models are needed to determine and control the in-network caching process (e.g., who can cache what content). Another direction of future research may be related to methods that will allow content encrypted by its producer to be cached by any node in the network. This content can be retrieved from an in-network cache by consumers (if the consumers learn the content name and send a request for it). However, consumers will not be able to decrypt the content, unless they have proper decryption keys issued by the producer. Such a mechanism ensures that only authorized users will have access to the encrypted content. However, it may be challenging to perform expensive cryptographic operations, given that applications might run on battery-powered devices.

\subsection{Objective Attacks}
In this subsection, we identify vulnerabilities that are amplified in wireless ICN-based environments.

\vspace{0.2cm}
\textbf{Passive Monitoring:}
Using content names to request and deliver data makes it trivial for nodes within a consumer's communication range to keep track of the requested content. Other factors, such as how often a consumer requests content or how much content is received, can reveal further information about a consumer. In a wired ICN system, the attacker has to compromise an on-path forwarder to obtain such information. Although not impossible, it creates a barrier for the intruder before it can obtain the information. In wireless environments, this information becomes widely available.

Encrypting packets (including their names) may be a direction that could help. However, this may make Interest forwarding infeasible, since intermediate nodes cannot process Interests based on their names unless they have a way to decrypt the names. Obfuscating the name of a packet (unless there is a real need for it, such as in censorship scenarios) defeats the objective of NDN to scale content distribution. Having Interests with name prefixes in plaintext and encrypted suffixes may achieve a proper tradeoff, which allows for Interest forwarding. However, its impact on the scalability of content distribution requires further investigation. In either case, a consumer will need to fetch and use the producer's public key for encryption or agree upon and exchange with the producer a symmetric encryption key.

\begin{figure}[t]
	\centering
	\includegraphics[width=\linewidth]{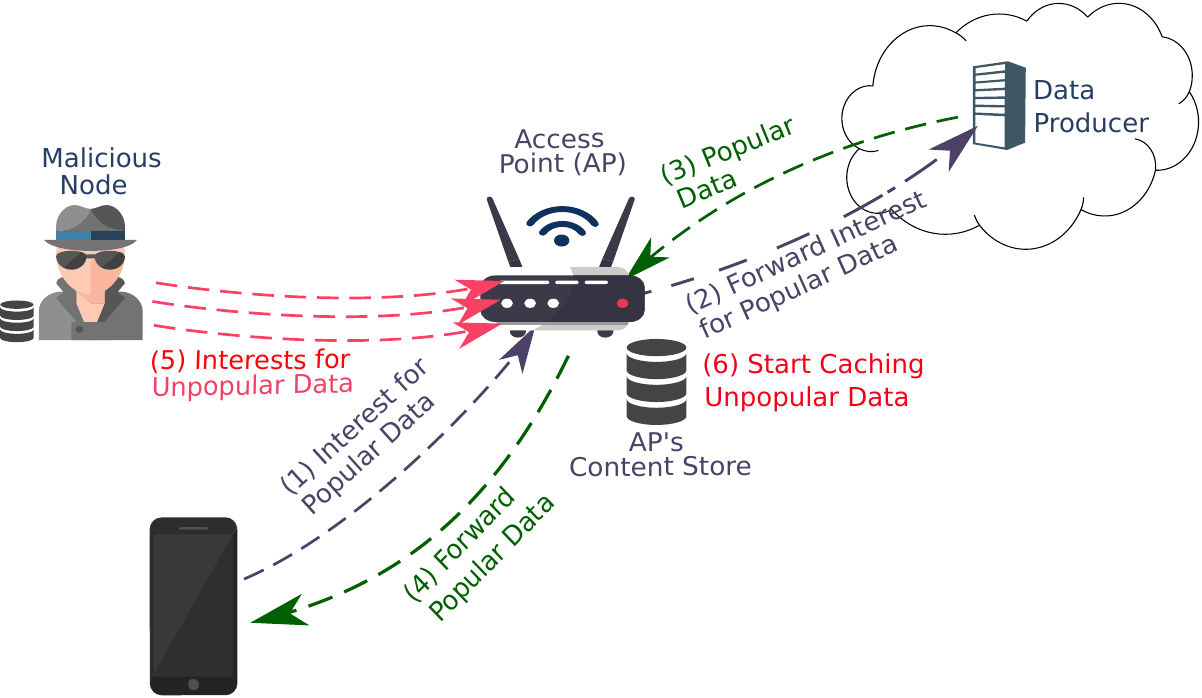}
	\caption{An example of a cache pollution attack.}
	\label{fig:cache_pollution}
\end{figure}

\vspace{0.2cm}
\textbf{Cache Manipulation at the Edge:}
In NDN, frequently requested content can be cached close to consumers. This may reduce the content retrieval latency and the network load. In a wireless environment, APs or edge routers become prime candidates for in-network caching. However, it is challenging to determine what content to cache. The simplest way is based on popularity, which from the perspective of an edge router can be determined in two ways:

\begin{itemize}[leftmargin=*]
	\item \emph{Interest based:} To make caching decisions, edge routers in infrastructure-based scenarios determine if the number of Interest packets for a specific content name exceeds a certain threshold within a given time interval. The challenge here is related to the fact that malicious nodes may generate bogus Interests for valid names, skewing the popularity results and forcing routers to cache content of their liking. To tackle this challenge, one approach might be for nodes to send signed Interests, so that receivers can verify the signature before sending the content back. However, this approach requires receiver nodes to retrieve the signer's certificate and verify its identity based on common trust anchors that have been established in advance~\cite{ramani2020rapid}. In disconnected environments, retrieving an entity's certificate and establishing trust anchors may be a challenge on its own.
	
	\item \emph {Popularity index:} This can be a variable contained in the Data packet by producers. Higher values will prompt edge routers to cache associated packets. However, for a producer to decide on such values, it needs to know how many Interests are sent by consumers for each packet, which is a challenge due to the pervasive NDN in-network caching. Moreover, competing producers could embed higher values into the Data packets to promote their own data. 
\end{itemize}

A malicious node may change the cached content's distribution, so that non-popular or even invalid content is cached in the content store. Figure~\ref{fig:cache_pollution} illustrates a cache pollution attack, where a malicious node changes the cache distribution of the edge by requesting non-popular content. After several requests for unpopular content, the edge node naturally assumes that the content is popular and caches it. To eliminate unpopular cached content, the content store can periodically verify the content popularity or cache content with a certain popularity as defined by the producer. However, the first method incurs computation overhead, while in the second method, the original producer may provide fake popularity values. Furthermore, a malicious node may generate fake content, which can then be cached by other nodes. Unless a strong verification method is employed for cached content, a few malicious nodes can create massive amounts of fake content in the network. 

\vspace{0.2cm}
\textbf{Flooding (resulting in Denial of Service):} A handful of attackers can overwhelm the network. In wireless ICN, an Interest may be propagated to direct neighbors and, subsequently, to more distant neighbors. Since there is no knowledge of valid or invalid content names at forwarding nodes, it is possible to generate Interests with random (flat or hierarchical) names. If a handful of nodes collide (on purpose or accidentally) while transmitting their Interests, a wireless network may get jammed, thus denying service to legitimate users.
It is important to keep the wireless communication models in perspective, as they create a trade-off. If nodes communicate through an AP, the AP becomes a bottleneck and the advantage of getting content from the closest neighbor is lost. However, this might alleviate the flooding problem described here. On the other hand, a multi-hop network may allow flexible content retrieval but is open to flooding attacks. 
An approach to detect Interest flooding attacks based on cumulative entropy has been proposed~\cite{xin2016novel}. The main idea of this design is identifying abnormal distributions of content requests and malicious name prefixes through relative entropy theory.

\begin{figure}[t]
	\centering
	\includegraphics[width=\linewidth]{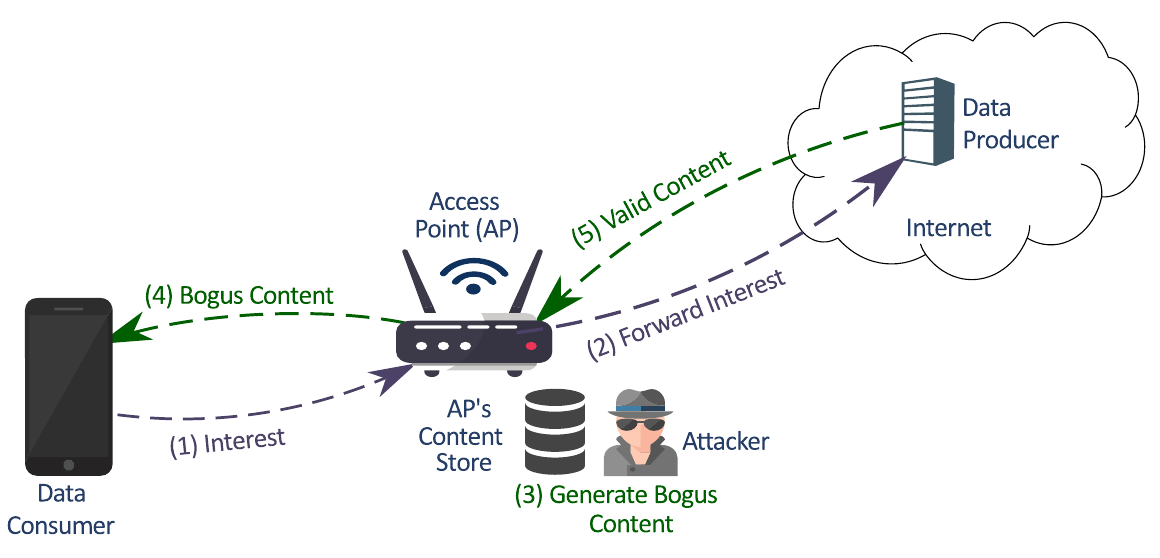}
	\caption{An example of a content poisoning attack.}
	\label{fig:content_poisoning}
\end{figure}

\vspace{0.2cm}
\textbf{Content Poisoning:} This attack aims to fill the content store of an intermediate node with invalid content, so that caching becomes worthless, and all requests must be forwarded upstream. An attacker must take control of an AP and inject bogus content to the network using valid names for the Interests to be satisfied.
Figure~\ref{fig:content_poisoning} illustrates such an attack scenario. Signature verification may be used to solve this issue, where a forwarder node verifies the signature of each Data packet before forwarding it. However, verifying the signature of each Data packet is impractical, thus a scheme to minimize unnecessary verifications and favor already-verified content in the content store may be a promising direction.

\vspace{0.2cm}
\textbf{Trust Relationships:}
Compared to IP-based networks, intermediate/forwarder nodes are involved in ICN, being able to respond to requests without getting permission from the content producer or letting the producer know. This transparent in-network caching makes trust among different network entities challenging and a critical aspect. Defining trust relationships between data producers, content stores, and content requesters is a direction that requires further investigation.
 
\subsection{Network and Devices}
Traditional wireless networks are connected in a binding manner with Service Set Identifier (SSID) and IP address assignments. However, NDN may result in evolving the MAC layer design. In an NDN-based network, where there may be no SSID bindings, it will be challenging to manage device joining and leaving, unless APs use certificates and verifications before processing any operation. Such potentially unrestricted devices in the network open up several challenges.

\vspace{0.2cm}
\textbf{Node Cooperation (Interests or Data packets):}
Nodes may selectively forward certain Interests or content. This may be done by both legitimate and malicious nodes. Unlike a wired switch or router, which is managed and follows the same protocol for all packets (unless compromised), a wireless device may be any handheld device. Depending on their cooperation level, wireless devices may choose to forward only those Interests, which result in a lower bandwidth or battery consumption. As data follows the reverse path of Interests, intermediate nodes may drop the returning data for certain Interests. Ensuring the same level of cooperation among devices and incentive mechanisms for participation will be a critical research direction.

\vspace{0.2cm}
\textbf{Resource Constrained Physical Layer:}
In terms of physical layer communication, data is transmitted in channels or slots by multiple devices in a shared wireless medium. The available bandwidth and channel capacity are highly limited, while their efficient and fair use is required. In existing systems, the network layer (IP) is responsible for slicing and fragmenting data, so that it can fit in standard-sized frames. Assuming that the physical layer technologies for NDN are the same as those of TCP/IP, they will carry over the limitations of the Maximum Transmission Unit (MTU). NDN is yet to finalize its fragmentation and reassembly mechanisms, however, the principle of flow balance (i.e., one Interest can bring back one Data packet) may limit the fragmentation possibilities. 

Fragmentation at the application layer, which may result in sending Interests for smaller pre-determined chunks of large content may be an interesting direction. Similarly, solutions for fragmentation and reassembly of content at the data link layer have been explored by the community~\cite{mosko2016icn, mosko2016ccnx, gundogan2018icn} and may also be an interesting direction to explore further. In either case, a malicious node could selectively request content, knowing that it will dominate the channel usage time, thus raising issues related to bandwidth utilization and fairness in a multi-user environment. Traditional fairness approaches at the transport layer are based on streams of packets (e.g., in TCP), however, in NDN, no such packet streams exist. Implementing fairness at layers below NDN requires considerable effort, especially in the wireless domain. Proposals for multiuser Orthogonal Frequency-Division Multiplexing (OFDM) fairness exist, but their use in a multi-hop NDN environment requires further exploration.

\subsection{Wireless ICN Devices}
Unlike the host-centric communication model, ICN nodes may be involved in both content discovery and delivery. They provide transparent in-network caching without informing content producers or consumers. However, caching may be exploited by malicious nodes to launch different attacks.

\begin{figure}[!t]
	\centering
	\includegraphics[width=\linewidth]{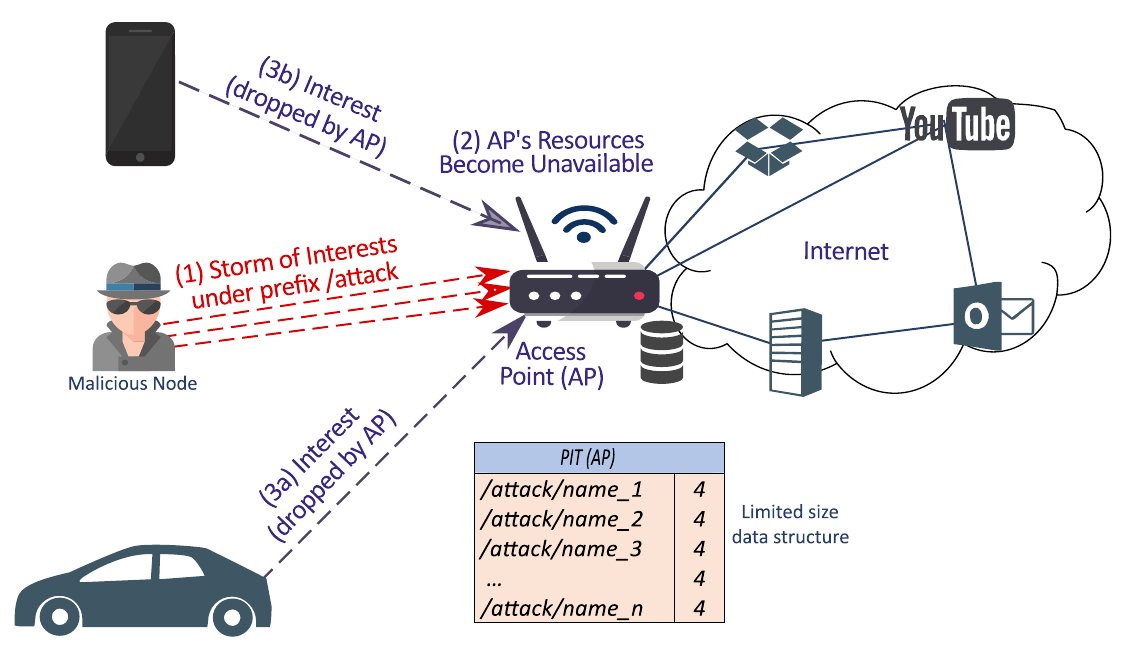}
	\caption{An example of a denial of service attack.}
	\label{fig:dos_attack}
\end{figure}

\vspace{0.2cm}
\textbf{Consumers:}
A consumer needs to know only the content name to send an Interest packet. Hence, it can send a storm of Interest packets asking either for available or unavailable content. This attack is known as Denial of Service, as depicted in Figure~\ref{fig:dos_attack}. If the content is available (valid content name), then the intermediate nodes may forward the request until reaching the producer or a replica node. In such a case, the attack's objective is to decrease the service performance and content availability and increase the response time. However, if the content is unavailable (invalid content name), the objective is attacking the network infrastructure itself.

Different countermeasures can be applied to solve this issue such as: (a) \textit{Per-Face Monitoring}: Collect and analyze Interest-related information (e.g., time-outs, ratio of incoming to satisfied Interests) to detect malicious Interests and nodes, (b) \textit{PIT Size Monitoring}: Monitor the size of PIT and whenever this reaches a threshold, the node starts a mitigation phase, and (c) \textit{Statistical Approaches}: Use statistical information of PIT and interfaces to identify abnormal Interests. Although these countermeasures may be effective, they can also block legitimate Interests. Such countermeasures have been utilized by the community as the building blocks of existing ICN frameworks, such as a framework to manage the Quality of Service (QoS) for ICN-based Internet of Things (IoT)~\cite{gundougan2020impact} and a framework to control (shape) the Interest traffic~\cite{wang2013improved}. Finally, frameworks for secure network measurements in NDN~\cite{nichols2019lessons} can facilitate network management operations and extract various measurements and statistics from forwarders to identify attacks, offer insights, and defend against these attacks.

\vspace{0.2cm}
\textbf{Intermediate Forwarders:}
The content name is the pillar of ICN content discovery in the sense that intermediate nodes use it to discover the nearest content copy. However, malicious forwarders may filter and block a set of names (e.g., drop all Interests for certain name patterns). Similarly, malicious forwarders can mistreat Data packets based on their names, by modifying the content during transmission (e.g., change the content payload) or by generating content on behalf of the legitimate producers (e.g., impersonation attack). Although each Data packet is signed by the original producer, and thus consumers may reject the altered data, consumers may not be able to access the valid content if the AP is the malicious node. A trust model is required among different network entities, including those who can cache the content, while consumers should be able to request content from the original producer skipping content stores. Smart contracts may be a suitable solution for such scenarios. However, realizing smart contracts in ICN is an open research question.

\vspace{0.2cm}
\textbf{Listener Nodes:}
All the wireless nodes within the communication range of a sender may receive both Interest and Data packets due to the broadcast nature of the wireless medium. Hence, a node can easily spoof routing and forwarding information and apply time analysis attacks in the network. It can further monitor the exchanged packets, identify popular requested content, and consequently request such content to violate user and data privacy. Under the same circumstances, an attacker may request unpopular content and force intermediate content stores to cache it, as if it was popular content.

\section{Conclusion and Future Research Directions}
\label{sec:conclusion}
This paper explored the security challenges and issues that may arise by deploying NDN/ICN in wireless environments from different points of view. Since NDN is still in its early design stages, these security issues should be handled as part of the core architecture design, instead of introducing add-on protocols. We believe that providing such add-ons will make NDN a complex architecture, as TCP/IP has also become.

\vspace{0.2cm}
\textbf{Future Research Directions:} As identified through our work, there is a number of directions for future research that deserve further investigation:

\begin{itemize}[leftmargin=*]

\item Exploration of content-centric MAC layer protocols in addition to mechanisms that allow existing MAC layer protocols to serve the needs of NDN/ICN.

\item Trust relationships are the basis for authenticating not only content producers in NDN/ICN, but every entity in a network. Establishing trust relationships in NDN/ICN wireless, and especially infrastructure-less, environments is challenging and approaches to facilitate that should be investigated.

\item Content encryption can be an effective solution to unauthorized content access. Attackers may be aware of the content name, being able to retrieve the content from an in-network cache. However, they will not be able to decrypt it, unless they have access to the corresponding decryption key(s). To this end, mechanisms for key distribution and revocation in wireless NDN/ICN environments should be further explored.
\end{itemize}

\section*{Acknowledgements}
This work was partially supported by the National Institutes of Health (NIGMS/P20GM109090), the National Science Foundation under award CNS-2016714, and the Nebraska University Collaboration Initiative.

\bibliographystyle{IEEEtran}

\begin{thebibliography}{15}
	\bibitem{xylomenos2014survey}
	G.~Xylomenos, C.~N. Ververidis, V.~A. Siris, N.~Fotiou, C.~Tsilopoulos,
	X.~Vasilakos, K.~V. Katsaros, and G.~C. Polyzos, ``{A survey of
		information-centric networking research},'' \emph{IEEE Communications Surveys
		\& Tutorials}, vol.~16, no.~2, pp. 1024--1049, 2014.
	
	\bibitem{yu2018content}
	Y.~Yu, Y.~Li, X.~Du, R.~Chen, and B.~Yang, ``{Content Protection in Named Data
		Networking: Challenges and Potential Solutions},'' \emph{IEEE Communications
		Magazine}, vol.~56, no.~11, pp. 82--87, 2018.
	
	\bibitem{zhang2018overview}
	Z.~Zhang, Y.~Yu, H.~Zhang, E.~Newberry, S.~Mastorakis, Y.~Li, A.~Afanasyev, and
	L.~Zhang, ``{An overview of security support in Named Data Networking},''
	\emph{IEEE Communications Magazine}, vol.~56, no.~11, pp. 62--68, 2018.
	
	\bibitem{zhang2014named}
	L.~Zhang, A.~Afanasyev, J.~Burke, V.~Jacobson, K.~Claffy, P.~Crowley,
	C.~Papadopoulos, L.~Wang, and B.~Zhang, ``{Named data networking},''
	\emph{ACM SIGCOMM Computer Communication Review}, vol.~44, no.~3, pp. 66--73,
	2014.
	
	\bibitem{kietzmann2017need}
	P.~Kietzmann, C.~G{\"u}ndo{\u{g}}an, T.~C. Schmidt, O.~Hahm, and
	M.~W{\"a}hlisch, ``The need for a name to mac address mapping in ndn: towards
	quantifying the resource gain,'' in \emph{ACM Conference on
		Information-Centric Networking}, 2017, pp. 36--42.
	
	\bibitem{sivaraman2020optimal}
	V.~Sivaraman, D.~Guha, and B.~Sikdar, ``{Optimal Pending Interest Table Size
		for ICN With Mobile Producers},'' \emph{IEEE/ACM Transactions on Networking},
	2020.
	
	\bibitem{khelifi2020blockchain}
	H.~Khelifi, S.~Luo, B.~Nour, H.~Moungla, S.~Ahmed, and M.~Guizani, ``{A
		Blockchain-based Architecture for Secure Vehicular Named Data Networks},''
	\emph{Computers \& Electrical Engineering}, vol.~86, p. 106715, 2020.
	
	\bibitem{ramani2020rapid}
	S.~K. Ramani and A.~Afanasyev, ``{Rapid Establishment of Transient Trust for
		NDN-Based Vehicular Networks},'' in \emph{IEEE International Conference on
		Communications Workshops (ICC Workshops)}.\hskip 1em plus 0.5em minus
	0.4em\relax IEEE, 2020, pp. 1--6.
	
	\bibitem{xin2016novel}
	Y.~Xin, Y.~Li, W.~Wang, W.~Li, and X.~Chen, ``{A novel interest flooding
		attacks detection and countermeasure scheme in NDN},'' in \emph{IEEE Global
		Communications Conference}, 2016, pp. 1--7.
	
	\bibitem{mosko2016icn}
	M.~Mosko and C.~Tschudin, ``{ICN’Begin-End’Hop by Hop Fragmentation},''
	\emph{IETF, Internet-Draft-work in progress}, 2016.
	
	\bibitem{mosko2016ccnx}
	M.~Mosko, ``{CCNx end-to-end fragmentation},'' \emph{IRTF Draft, Palo Alto
		Research Center, Inc}, 2016.
	
	\bibitem{gundogan2018icn}
	C.~G{\"u}ndogan, T.~C. Schmidt, M.~W{\"a}hlisch, C.~Scherb, C.~Marxer, and
	C.~Tschudin, ``{ICN Adaptation to LowPAN Networks (ICN LoWPAN)},'' \emph{IRTF
		Internet Draft--work in progress 02}, 2018.
	
	\bibitem{gundougan2020impact}
	C.~G{\"u}ndo{\u{g}}an, J.~Pfender, P.~Kietzmann, T.~C. Schmidt, and
	M.~W{\"a}hlisch, ``{On the impact of QoS management in an Information-centric
		Internet of Things},'' \emph{Computer Communications}, 2020.
	
	\bibitem{wang2013improved}
	Y.~Wang, N.~Rozhnova, A.~Narayanan, D.~Oran, and I.~Rhee, ``{An improved
		hop-by-hop interest shaper for congestion control in named data
		networking},'' \emph{ACM SIGCOMM Computer Communication Review}, vol.~43,
	no.~4, pp. 55--60, 2013.
	
	\bibitem{nichols2019lessons}
	K.~Nichols, ``{Lessons learned building a secure network measurement framework
		using basic NDN},'' in \emph{ACM Conference on Information-Centric
		Networking}, 2019, pp. 112--122.
\end{thebibliography}

\end{document}